\documentclass[a4paper]{jpconf}

\usepackage{amssymb}
\usepackage{amsmath}
\usepackage{afterpage}

\newcommand{\bea}{\begin{eqnarray}}
\newcommand{\eea}{\end{eqnarray}}
\newcommand{\be}{\begin{equation}}
\newcommand{\ee}{\end{equation}}
\newcommand{\no}{{\nonumber}}
\renewcommand{\eref}[1]{Eq.~\eqref{#1}}
\newcommand{\secref}[1]{Sec.~\ref{#1}}

\newcommand{\tabref}[1]{Table~\ref{#1}}
\renewcommand{\tr}{\mathop{\mathrm{Tr}}}

\newcommand{\slD}{\mbox{$D$ \hspace{-1.3em} $\slash$}}

\begin{document}

\title{Zero temperature properties of mesons and baryons from an
  extended linear sigma-model}
\author{P Kov{\'a}cs and Gy Wolf}
\address{Institute for Particle and Nuclear Physics, Wigner Research
  Centre for Physics, Hungarian Academy of Sciences, XII. Konkoly
  Thege Mikl{\'o}s {\'u}t 29-33, 1121 Budapest, Hungary}
\ead{kovacs.peter@wigner.mta.hu}

\begin{abstract}
  An extended linear sigma model with mesons ($q\bar{q}$ states) and
  baryons ($qqq$ states) is presented. The model contains a low energy
  multiplet for every hadronic particle type, namely a scalar, a
  pseudoscalar, a vector and an axialvector nonet, a baryon octet and a
  baryon decuplet. The model parameters are determined through
  a multiparametric minimalization with the help of well known physical
  quantities. It is found that the considered zero temperature
  quantities (masses and decay widths) can be described well at
  tree-level and are in good agreement with the experimental data.
\end{abstract}

\section{Introduction}

The vacuum properties of strong interaction are very hard to investigate
within the framework of QCD -- the fundamental theory of strong
interaction -- (see e.g.\cite{textbook:Peskin}), which is due to its subtlety at low
energies. Consequently, instead of solving QCD one can set up an
effective theory, which reflects some properties of the original theory.   
The underlying principle in the construction of such theories is that
they share the same global symmetries as QCD.

For zero masses of the $u$, $d$ and $s$ quarks, the global symmetry of
QCD is $U(3)_R \times U(3)_L$, the so-called chiral symmetry. In the
vacuum, this chiral symmetry is spontaneously broken due to the
existence of a quark-antiquark condensate. The
chiral symmetry can be realized in different ways, nonlinearly
\cite{nonlin_sigma} and linearly \cite{Gasiorowicz:1969}, which we
choose here. Accordingly, in this paper we set up an extended linear
sigma model, which contains mesonic and baryonic degrees of freedom. 
The previous versions of our model \cite{elsm_su2, elsm_2013} contained the scalar,
pseudoscalar, vector- and axial-vector nonets. The vacuum
phenomenology of mesons was described very well in that model. As an
improvement, in this paper we include additionally the nucleon-octet
and the Delta-decuplet to extend the vacuum phenomenology for baryons
as well. Another approach to baryon phenomenology can be found in
\cite{Zhuang_1999}. 

Our paper is organized as follows. In the \secref{Sec:model} we
briefly present the model, while in \secref{Sec:quantities} we
describe how to calculate various tree-level quantities. The
\secref{Sec:result} is dedicated to the results in the mesonic and
baryonic sector and finally we conclude in \secref{Sec:conclusion}.

\section{The Model}
\label{Sec:model}

The model can be defined through a Lagrangian consisting of a purely
mesonic and a baryonic-mesonic part as $\mathcal{L} =
\mathcal{L}_{\text{meson}} + \mathcal{L}_{\text{baryon}}$. The terms
in the meson part are limited by chiral and dilaton symmetry (for
details see \cite{elsm_2013}), while in case of the baryon part we
included all the $SU(3)_V$ invariants which can produce baryon masses
-- with different masses for different particles in the given multiplet --
and decuplet decays with the lowest possible dimension ($B-B-\Phi-\Phi$,
$\Delta-\Delta-\Phi-\Phi$ and $\Delta-B-\Phi$ terms).
The mesonic part has the following form
\begin{align}
\mathcal{L_{\text{meson}}}  &
= \tr [(D_{\mu}\Phi)^{\dagger}(D_{\mu}\Phi)]-m_{0}
^{2}\tr(\Phi^{\dagger}\Phi)-\lambda_{1}
[\tr(\Phi^{\dagger}\Phi)]^{2}-\lambda_{2}
\tr(\Phi^{\dagger}\Phi)^{2}{\nonumber}\\
&  -\frac{1}{4}\tr(L_{\mu\nu}^{2}+R_{\mu\nu}^{2}
)+\tr\left[  \left(  \frac{m_{1}^{2}}{2}+\Delta\right)
(L_{\mu}^{2}+R_{\mu}^{2})\right]
+\tr[H(\Phi+\Phi^{\dagger})]{\nonumber}\\
&  +c_{1}(\det\Phi-\det\Phi^{\dagger})^{2}+i\frac{g_{2}}{2}
(\tr\{L_{\mu\nu}[L^{\mu},L^{\nu}
]\}+\tr\{R_{\mu\nu}[R^{\mu},R^{\nu}]\}){\nonumber}\\
&  +\frac{h_{1}}{2}\tr(\Phi^{\dagger}\Phi
)\tr(L_{\mu}^{2}+R_{\mu}^{2})+h_{2}
\tr[(L_{\mu}\Phi)^{2}+(\Phi R_{\mu} )^{2}]+2h_{3}
\tr(L_{\mu}\Phi R^{\mu}\Phi^{\dagger}).{\nonumber}\\
&  +g_{3}[\tr(L_{\mu}L_{\nu}L^{\mu}L^{\nu}
)+\tr(R_{\mu}R_{\nu}R^{\mu}R^{\nu})]+g_{4}
[\tr\left(  L_{\mu}L^{\mu}L_{\nu}L^{\nu}\right)
+\tr\left(  R_{\mu}R^{\mu}R_{\nu}R^{\nu}\right)
]{\nonumber}\\
&  +g_{5}\tr\left(  L_{\mu}L^{\mu}\right)
\,\tr\left(  R_{\nu}R^{\nu}\right)  +g_{6}
[\tr(L_{\mu}L^{\mu})\,\tr(L_{\nu}L^{\nu
})+\tr(R_{\mu}R^{\mu})\,\tr(R_{\nu}R^{\nu
})]\text{  ,} \label{eq:Lagrangian}
\end{align}
where
\begin{align}
D^{\mu}\Phi &  \equiv\partial^{\mu}\Phi-ig_{1}(L^{\mu}\Phi-\Phi R^{\mu
})-ieA^{e\,\mu}[T_{3},\Phi]\;,\nonumber\\
L^{\mu\nu}  &  \equiv\partial^{\mu}L^{\nu}-ieA^{e\,\mu}[T_{3},L^{\nu}]-\left\{
\partial^{\nu}L^{\mu}-ieA^{e\,\nu}[T_{3},L^{\mu}]\right\}\;  ,\nonumber\\
R^{\mu\nu}  &  \equiv\partial^{\mu}R^{\nu}-ieA^{e\,\mu}[T_{3},R^{\nu}]-\left\{
\partial^{\nu}R^{\mu}-ieA^{e\,\nu}[T_{3},R^{\mu}]\right\}
\; ,\nonumber
\end{align}
The quantities $\Phi = \sum_{i=0}^{8}(S_{i}+iP_{i})T_{i}$,
$L^{\mu}/R^{\mu} = \sum_{i=0}^{8}(V_{i}^{\mu} \pm A_{i}^{\mu})T_{i} $
represent the scalar-pseudoscalar nonets and the left-/right-handed vector
nonets. $T_{i}\,(i=0,\ldots,8)$ denote the generators of $U(3)$, while
$S_{i}$ represents the scalar, $P_{i}$ the pseudoscalar, $V_{i}^{\mu}$ the vector, and
$A_{i}^{\mu}$ the axial-vector meson fields, and $A^{e\,\mu}$ is the
electromagnetic field. $H$ and $\Delta$ are some constant
external fields. It should be noted that in the $(0-8)$
sector\footnote{A 2 by 2 segment of the mass matrix consisting of the
components: $00, 08, 80, 88$}
of the scalars and pseudoscalars  there is a mixing and it is more
suitable to use the non strange -- strange basis defined as
$\varphi_{N} = 1/\sqrt{3}(\sqrt{2} \varphi_{0} + \varphi_{8})$,
$\varphi_{S} = 1/\sqrt{3}(\varphi_{0} - \sqrt{2} \varphi_{8} )$ for
$\varphi_{i} \in (S_i,P_i,V_i^{\mu},A_i^{\mu})$. 

Moving on to the baryonic-mesonic part, the Lagrangian is given by
\begin{align}
  \mathcal{L}_{\text{baryon}} & = \tr \left[\bar{B}\left(i \slD-M_{(8)}\right)B\right] \no\\
  & -  \tr\left\{\bar{\Delta}_{\mu}\cdot \left[ \left(i \slD -
        M_{(10)}\right)g^{\mu\nu} - i\left(\gamma^{\mu}D^{\nu} + \gamma^{\nu}D^{\mu}\right)+ 
      \gamma^{\mu}\left(i \slD + M_{(10)}\right)\gamma^{\nu}\right]\Delta_{\nu}\right\} \no\\
  & +  C\tr\left[\bar{\Delta}^{\mu}\cdot\left(
      -\frac{1}{f}(\partial_{\mu} - ieA^{e}_{\mu}[T_3,\Phi]) -
      \frac{1}{f}[\Phi,V_{\mu}] + A_{\mu} \right)B\right] +
  \text{h. c.}  \no \\
  & -  \xi_1^{} \tr \left(\bar{B}B\right ) \tr \left(
    \Phi^{\dagger}\Phi\right) - \xi_2^{} \tr \left( \bar{B} \{\{\Phi,
    \Phi^{\dagger}\}, B \} \right) - \xi_3^{} \tr \left( \bar{B}
    [\{\Phi,\Phi^{\dagger}\},B] \right)\no\\
  & -  \xi_4^{} \left(\tr \left(\bar{B}\Phi\right ) \tr \left(
    \Phi^{\dagger}B\right) + \tr \left(\bar{B}\Phi^{\dagger}\right ) \tr \left(
    \Phi B\right) \right) -  \xi_5^{} \tr \left(\bar{B} \{ [\Phi,
\Phi^{\dagger}], B \} \right) \label{eq:lagr_baryon}\\
  & -  \xi_6^{} \tr \left(\bar{B} [[\Phi,\Phi^{\dagger}], B] \right) -
  \xi_7^{} \left(\tr \left(\bar{B}\Phi\right ) \tr \left( \Phi^{\dagger}B\right) - \tr
    \left(\bar{B}\Phi^{\dagger}\right ) \tr \left( \Phi B\right) \right)\no \\
  & -  \xi_8^{} \left(\tr \left(\bar{B}\Phi B \Phi^{\dagger} \right ) -
    \tr \left(\bar{B}\Phi^{\dagger} B \Phi \right) \right) + \chi_1^{}
    \tr \left( \bar{\Delta} \cdot \Delta \right) \tr \left( \Phi^{\dagger}\Phi\right)\no \\
  & +  \chi_2^{} \tr \left( (\bar{\Delta} \cdot \Delta)
    \{\Phi,\Phi^{\dagger}\} \right) + \chi_3^{} \tr \left( (\bar{\Delta} \cdot
    \Phi) (\Phi^{\dagger} \cdot \Delta) + (\bar{\Delta} \cdot
    \Phi^{\dagger}) (\Phi \cdot \Delta) \right) \no\\
  & +  \chi_4^{} \tr \left( (\bar{\Delta} \cdot \Delta) [\Phi,\Phi^{\dagger}] \right), \no
\end{align}
where $B = \sqrt{2}\sum_{i=1}^{8}B_{a}T_{a}$ and $\Delta_{\mu}$ stands
for the baryon octet and decuplet. $M_{(8)}$ and $M_{(10)}$ are the
bare masses of the baryon octet and decuplet. $f$ is the pion decay
constant, while  $[\;,\;]$ and $\{\; ,\;\}$ denote the commutator and
the anticommutator. Here the meson-baryon interaction terms are all
the possible $SU(3)_V$ invariants that can be written down with the
given number of fields \cite{SU3_inv, Semke_thesis}. The covariant
derivatives are defined as 
\begin{align}
  D_{\mu} B & = \partial_{\mu} B + i [B,V_{\mu}] + \frac{1}{f}\left\{
    [A_{\mu},\Phi],B\right\}, \no \\
  D_{\mu} \Delta_{\nu}^{ijk}& = \partial \Delta_{\nu}^{ijk} +
  \left(\frac{1}{f} [A_{\mu},\Phi]^{i}_{l} - iV_{\mu\,l}^{\
      \,i}\right) \Delta_{\nu}^{ljk} + \left(\frac{1}{f}
    [A_{\mu},\Phi]^{j}_{l} - iV_{\mu\,l}^{\ \,j}\right)
  \Delta_{\nu}^{ilk} + \left(\frac{1}{f} [A_{\mu},\Phi]^{k}_{l} -
    iV_{\mu\,l}^{\ \,k}\right) \Delta_{\nu}^{ijl}, \no 
\end{align}
and the following dot notation is used:
\be
  (\bar{\Delta} \cdot \Delta)_{k}^{m} \equiv
  \bar{\Delta}_{ijk}\Delta^{ijm}, \quad (\bar{\Delta} \cdot
  \Phi)_{k}^{m} \equiv \bar{\Delta}_{ijk} \Phi_{l}^{i}\epsilon^{jlm},
  \quad (\Phi \cdot \Delta)_{k}^{m} \equiv \Delta^{ijm}
  \Phi_{i}^{l}\epsilon_{jlm}. 
\ee
From the given Lagrangian various tree-level quantities such as masses
and decay widths can be calculated and can be used to determine the
unknown parameters of the model. During this parametrization process
we can check how well the physical spectrum is reproduced. In the next
section calculation of tree-level quantities and the parametrization
are discussed.

\section{Tree-level quantities and parametrization}
\label{Sec:quantities}

As a standard procedure in a spontaneously broken theory we assume
non-zero vacuum expectation values (vev) to certain fields, in our
case to the $\sigma_N \equiv 1/\sqrt{3}(\sqrt{2} \sigma_{0} +
\sigma_{8})$ and $\sigma_S \equiv 1/\sqrt{3}(\sigma_{0} - \sqrt{2}
\sigma_{8} )$ scalar fields, and denote their vev by $\phi_N$ and
$\phi_S$. After that the $\sigma_N$ and $\sigma_S$ fields are shifted
with their non zero vev's $\phi_N$ and $\phi_S$. Consequently, the
quadratic and three-coupling terms of the Lagrangian can be determined
from which the masses and the decay widths originate. However, it
should be noted that as a technical difficulty -- due to the
$\sigma_N$ and $\sigma_S$ field shifts -- different particle mixings
emerge. In detail, there will be mixings in the $N-S$ (or $0-8$)
sector of the scalar and pseudoscalar octets and between the
vector-scalar and axialvector-pseudoscalar nonets. The $N-S$ mixings
can be resolved by some orthogonal transformation, while the other
mixings by redefinition of certain (axial-)vector fields. The details
can be found in \cite{elsm_2013} together with explicit expressions
for the meson masses and various decay widths.

Regarding the baryon sector the tree-level octet and decuplet masses
-- from the terms of the Lagrangian quadratic in the fields $B$ and
$\Delta_{\mu}$ -- are found to be
\begin{align}
  m_p = m_n &= M_{(8)} + \frac{1}{2}\xi_2(\Phi_N^2 + 2\Phi_S^2) +
  \frac{1}{2}\xi_3 (\Phi_N^2 - 2\Phi_S^2),\nonumber \\
  m_{\Xi} &= M_{(8)} + \frac{1}{2}\xi_2(\Phi_N^2 + 2\Phi_S^2) -
  \frac{1}{2}\xi_3 (\Phi_N^2 - 2\Phi_S^2), \nonumber\\
  m_{\Sigma} &= M_{(8)} + \xi_2 \Phi_N^2, \nonumber\\
  m_{\Lambda} &= M_{(8)} + \frac{1}{3}\xi_2(\Phi_N^2 + 4\Phi_S^2) +
  \frac{1}{3}\xi_4 (\Phi_N - \sqrt{2}\Phi_S)^2,
%
  m_{\Delta} &= M_{(10)}  + \frac{1}{2} \chi_2^{} \Phi_N^2,\nonumber \\
  m_{\Sigma^{\star}} &= M_{(10)} + \frac{1}{3}\chi_{2}^{} (\Phi_N^2 +
  \Phi_S^2) + \frac{1}{6} \chi_{3}^{} (\Phi_N - \sqrt{2}\Phi_S)^2,\nonumber \\
  m_{\Xi^{\star}} &= M_{(10)} + \frac{1}{6} \chi_{2}^{} (\Phi_N^2 + 4
  \Phi_S^2) + \frac{1}{6} \chi_{3}^{} (\Phi_N - \sqrt{2}\Phi_S)^2,\nonumber \\
  m_{\Omega} &= M_{(10)} + \chi_{2}^{} \Phi_N^2.
\end{align}
Beside the masses one can consider two-body decays of the decuplet
baryons. According to PDG \cite{PDG} there are four such physically allowed
decays, 
\begin{equation}
\Delta \to p \pi, \quad \Sigma^{\star} \to \Lambda \pi,\quad
\Xi^{\star} \to \Xi \pi, \quad \Sigma^{\star} \to \Sigma \pi. \label{eq:Delta_decay}
\end{equation}
And the decay widths are given by
\begin{align}
  \Gamma_{\Delta \to \pi p} &= \frac{k^3_{\Delta}}{24 m_{\Delta}}
  (m_{p} + E_{p}) G^2,\quad 
  \Gamma_{\Sigma^{\star} \to \pi \Lambda} =
  \frac{k^3_{\Sigma^{\star}}}{48 m_{\Sigma^{\star}}} (m_{\Lambda} +
  E_{\Lambda})  G^2,\no \\
  \Gamma_{\Xi^{\star} \to \pi \Xi} &= \frac{k^3_{\Xi^{\star}}}{48
    m_{\Xi^{\star}}} (m_{\Xi} + E_{\Xi}) G^2,\quad
  \Gamma_{\Sigma^{\star} \to \pi \Sigma} =
  \frac{k^3_{\Sigma^{\star}}}{72 m_{\Sigma^{\star}}} (m_{\Sigma} +
  E_{\Sigma}) G^2,\label{Eq:decuplet_decay}
\end{align}
with
\be
G^2 = C^2Z_{\pi}^2\left(w_{a_1}^2+\frac{1}{f^2} \right).\no
\ee
As it can be seen from the Lagrangian there are 30 unknown parameters
of the model, 14 in the meson sector and 16 in the baryon sector.
However some of them can be set to zero without the loss of
generality, some of them not even appear in the formulas of the
physical quantities considered here, while some of them appear only in
certain combinations. All in all there is 19 parameters which should
be determined. These parameters are determined through the comparison
of the calculated tree-level expressions -- from which we have 23 --
of the spectrum and decay
widths with their experimental value taken from \cite{PDG} with
artificially increased errors ($5\%$ for the masses\footnote{The
  isospin violation in some cases (e.g. for pion) has the order of
  $5\%$} and $10\%$ for the decay widths, since we do not expect from
a tree-level model to be more precise). Our strategy is that first we
set the parameters of the
meson sector \cite{elsm_2013} and then we fit the remaining parameters
of the meson-baryon interaction terms. For this we used a
$\chi^2$-minimalization, which was realized with a
multiparametric minimalization code (MINUIT \cite{MINUIT}).

\section{Results}
\label{Sec:result}

Using the above mentioned $\chi^2$-minimalization method in the meson
sector we obtain results summarized in \tabref{meson}, which can also
be found in \cite{elsm_2013}.
\begin{table}[ht!]
\caption{\label{meson}Calculated and experimental values of meson observables}
\begin{center}
\lineup
\begin{tabular}{*{4}{l}}
\br
Observable & Fit [MeV] & PDG [MeV] & Error [MeV] \cr
\mr
$m_{\pi}$ & $141.0 \pm5.8$ & $137.3$ & $\pm6.9$ \cr
$m_{K}$ & $485.6 \pm3.0$ & $495.6$ & $\pm24.8$ \cr
$m_{\eta}$ & $509.4 \pm3.0$ & $547.9$ & $ \pm27.4$ \cr
$m_{\eta^{\prime}}$ & $962.5 \pm5.6$ & $957.8$ & $\pm47.9$ \cr 
$m_{\rho}$ & $783.1 \pm7.0$ & $775.5$ & $ \pm38.8$ \cr
$m_{K^{\star}}$ & $885.1 \pm6.3$ & $893.8$ & $ \pm44.7$ \cr
$m_{\phi}$  & $975.1 \pm6.4$ & $1019.5$ & $ \pm51.0$ \cr
$m_{a_{1}}$ & $1186 \pm6$ & $1230$ & $ \pm62$ \cr
$m_{f_{1}(1420)}$ & $1372.5 \pm5.3$ & $1426.4$ & $ \pm71.3$ \cr
$m_{a_{0}}$ & $1363 \pm1$ & $1474$ & $ \pm74$ \cr
$m_{K_{0}^{\star}}$ & $1450 \pm1$ & $1425$ & $ \pm71$ \cr
$\Gamma_{\rho\rightarrow\pi\pi}$ & $160.9 \pm4.4$ & $149.1$ & $ \pm7.4$ \cr
$\Gamma_{K^{\star}\rightarrow K\pi}$ & $44.6 \pm1.9$ & $46.2$ & $ \pm2.3$ \cr
$\Gamma_{\phi\rightarrow\bar{K}K}$  & $3.34 \pm0.14$  & $3.54$ & $ \pm0.18$ \cr
$\Gamma_{a_{1}\rightarrow\rho\pi}$  & $549 \pm43$ & $425$ & $ \pm175$ \cr
$\Gamma_{a_{1}\rightarrow\pi\gamma}$ & $0.66 \pm0.01$ & $0.64$ & $ \pm0.25$ \cr
$\Gamma_{f_{1}(1420)\rightarrow K^{\star}K}$ & $44.6 \pm39.9$ & $43.9$ & $\pm 2.2$ \cr
$\Gamma_{a_{0}}$ & $266 \pm12$ & $265$ & $ \pm13$ \cr
$\Gamma_{K_{0}^{\star}\rightarrow K\pi}$ & $285 \pm12$ & $270$ & $\pm80$ \cr
\br
\end{tabular}
\end{center}
\end{table}

In the baryon sector at first we only investigated the decuplet
decays. The results are given in \tabref{baryon}.
\begin{table}[ht!]
\caption{\label{baryon}Calculated and experimental values of baryon observables}
\begin{center}
\lineup
\begin{tabular}{*{4}{l}}
\br
Observable & Fit [MeV] & PDG [MeV] & Error [MeV] \cr
\mr
$\Gamma_{\Delta \rightarrow p \pi}$ & $67.3$  & $110.0 $ & $\pm 11.0$ \cr
$\Gamma_{\Sigma^{\star} \rightarrow \Lambda \pi}$ & $27.0$ & $32.0 $ & $\pm 3.2$ \cr 
$\Gamma_{\Sigma^{\star} \rightarrow \Sigma \pi}$ & $4.9$ & $4.3 $ & $\pm 0.4$ \cr
$\Gamma_{\Xi^{\star} \rightarrow \Xi \pi}$ & $11.2$ & $9.5 $ & $\pm 1.0$ \cr
\br
\end{tabular}
\end{center}
\end{table}

It can be seen that the meson observables are reproduced very well in
general, while the decuplet decays show a fair correspondence. It is
worth to note that in case of the the decuplet decays their tree-level
expressions \eref{Eq:decuplet_decay} only differ in their kinematic
parts and we have only one parameter to fit for the four observables,
which can cause the deviation from the experimental value. 

\section{Conclusion}
\label{Sec:conclusion}

We have presented an extended linear sigma model with meson and baryon
degrees of freedom. This is a possible extension with baryon octet and
decuplet of our previous meson model \cite{elsm_2013}.
We included interaction terms, such as $\Delta-B-P$ and
$B-B-\Phi-\Phi$ to describe $\Delta$ decays and baryon masses. 
We calculated the tree-level masses and physically relevant decuplet
decay widths and we found that in general they are in good agreement
with the experimental data taken from the PDG \cite{PDG}.

As a continuation we plan to include the baryon masses to the fit and
to add other (higher dimension) interaction terms containing
derivatives which are important in case of scattering processes
\cite{Zhuang_1999}. Our further aim is to go to finite
temperature and/or densities with all these fields included in our
model. 

\ack

P.\ Kov{\'a}cs and Gy.\ Wolf were partially supported by the Hungarian
OTKA funds NK101438 and K109462.

\end{document}